\documentclass[11pt]{article}

\usepackage[margin=1in]{geometry}
\usepackage{amsmath,amssymb}
\usepackage{booktabs}
\usepackage{graphicx}
\usepackage{hyperref}
\usepackage{url}
\usepackage{algorithm}
\usepackage{algorithmic}
\usepackage{xcolor}
\usepackage{enumitem}
\usepackage{pifont}
\usepackage{multirow}
\usepackage{tabularx}
\usepackage{microtype}
\hypersetup{hidelinks}

\newcommand{\cmark}{\ding{51}}
\newcommand{\xmark}{\ding{55}}

\title{\textbf{MLIPilot}: LLM-Driven Auto-Research for Machine-Learned Interatomic Potentials}
\author{Etinosa Osaro\thanks{\texttt{eosaro@psiquantum.com}}, Santosh Adhikari, Stamatia Zavitsanou,\\ Kelsey Parker,   and Dario Rocca\thanks{\texttt{drocca@psiquantum.com}}\\[4pt]
\textit{PsiQuantum, 700 Hansen Way, Palo Alto, CA 94304}}
\date{May 2026}

\begin{document}

\maketitle

\begin{abstract}
\noindent
Constructing production-quality machine-learned interatomic potentials
(MLIPs) requires balancing accuracy, dynamical stability, and computational throughput under constraints that are not captured by a single training loss.
We introduce \textbf{MLIPilot}, an auto-research framework in which
tool-calling large language models propose hypotheses, edit MLIP training
code, launch HPC jobs, and accept or revert changes using a fixed,
physically constrained scorecard.
We evaluate MLIPilot on MACE potential optimization using both commercial
and open-weight LLM agents, including GPT-5.5, GPT-4.1, Mistral-24B, and
Qwen3-32B. The benchmarks span molecular and periodic settings: a
QM7-derived dataset for which we generated B3LYP/6-31G(d) energies and
forces, and a Cu EMT dataset with periodic copper supercells labeled by
ASE's Effective Medium Theory calculator.
Across these benchmarks, the strongest agents move initially
constraint-violating baselines to accepted models by discovering useful
training strategies, including output normalization, loss-function changes,
progressive training schedules, and model-capacity adjustments. These
results suggest that LLM agents can serve as autonomous operators for
scientific machine-learning workflows when their search is constrained by
domain-specific validation criteria, shifting part of MLIP development from
manual trial-and-error toward auditable, automated experimentation.
\end{abstract}

\section{Introduction}

Machine-learned interatomic potentials approximate quantum-mechanical energy surfaces at a fraction of the cost of density functional theory, enabling nanosecond-scale molecular dynamics of $10^4$--$10^6$ atoms~\cite{batatia_mace_2022,batzner_nequip_2022}.
Yet developing a production MLIP remains an exercise in multi-objective constrained optimization: energy and force accuracy must satisfy application-specific thresholds; NVE dynamics must conserve energy over several picosecond trajectories; and inference throughput must support practical simulation timescales.
These objectives are coupled in non-obvious ways.
Aggressive energy-loss weighting can destabilize molecular dynamics.
Deeper networks improve accuracy but reduce throughput below usable limits.
Extended training may overfit in ways that manifest not as increased validation loss but as explosive NVE drift, a catastrophic failure mode invisible to standard metrics.

Human experts navigate this space through iterative experimentation guided by physical intuition, a slow and irreproducible process that has motivated decades of work on automated machine learning~\cite{hutter_automl_book_2019,he_automl_survey_2021}.

The question motivating this work is whether large language models~\cite{brown_gpt3_2020,openai_gpt4_2023,touvron_llama_2023}, given appropriate tools and feedback, can serve as autonomous interatomic-potential researchers.
Karpathy's AutoResearch framework\footnote{\url{https://github.com/karpathy/AutoResearch}}~\cite{karpathy_autoresearch,karpathy_autoresearch_session} demonstrated precisely this paradigm for general ML: an LLM agent iteratively edits training code within a fixed evaluation harness, submits experiments, and retains only improvements, reducing Time-to-GPT-2 from 2.02 to 1.8~hours across approximately 700 autonomous experiments on nanochat LLM pretraining.
The ``fix evaluation, vary training, keep improvements'' pattern maps naturally to MLIP development, with one critical modification: the scalar validation loss must be replaced by a multi-objective scorecard with hard physical-feasibility constraints that reject dynamically unstable models regardless of composite improvement.

We present three contributions:
\begin{enumerate}[nosep,leftmargin=*]
    \item \textbf{MLIPilot}: an agentic framework coupling tool-calling LLMs with Slurm HPC execution, SHA-256 integrity enforcement, and hypothesis-driven experiment logging.
    \item A \textbf{physically constrained scorecard} with targets parsed from natural-language prompts and hard gates at $4\times$ target that guarantee dynamical stability of accepted models.
    \item A \textbf{multi-agent benchmark} on two datasets revealing that scientific reasoning quality, not model scale or token budget, determines optimization success, and that open-weight models can outperform larger proprietary ones on specific tasks.
\end{enumerate}

\section{Related Work}

\textbf{Equivariant MLIPs.}
The field has progressed from invariant descriptors (BPNN~\cite{behler_bpnn_2007}, SchNet~\cite{schutt_schnet_2018}) through angular features (DimeNet++~\cite{gasteiger_dimenetpp_2020}) to fully E(3)-equivariant message passing (NequIP~\cite{batzner_nequip_2022}, Allegro~\cite{musaelian_allegro_2023}, PaiNN~\cite{schutt_painn_2021}, GemNet~\cite{gasteiger_gemnet_2021}).
MACE~\cite{batatia_mace_2022} combines higher-order equivariant messages with multi-body correlation, achieving favorable accuracy-speed Pareto fronts.
Universal potentials (MACE-MP~\cite{batatia_mace_mp_2024}, M3GNet~\cite{chen_m3gnet_2022}, CHGNet~\cite{deng_chgnet_2023}) cover the periodic table but require system-specific fine-tuning for quantitative accuracy~\cite{fu_forces_are_not_enough_2023}.
Our work is orthogonal: MLIPilot optimizes the \emph{training procedure} for any architecture.

\textbf{LLM agents for science.}
The emergence of tool-calling capabilities in LLMs~\cite{schick_toolformer_2023,qin_tool_learning_2023} has enabled autonomous scientific agents.
Coscientist~\cite{boiko_coscientist_2023} demonstrated closed-loop chemical synthesis with robotic execution; ChemCrow~\cite{bran_chemcrow_2024} augmented LLMs with 18 chemistry tools; Jablonka et al.~\cite{jablonka_llm_chemistry_2024} applied GPT-4 to materials data analysis.
In ML research automation, FunSearch~\cite{romera_paredes_funsearch_2024} used LLMs to discover novel mathematical algorithms, and the AI Scientist~\cite{lu_ai_scientist_2024} automated paper writing from idea generation through experimentation.
Karpathy's AutoResearch~\cite{karpathy_autoresearch} established the paradigm closest to our work: LLM-driven iterative code improvement within a fixed evaluation harness.
MLIPilot extends this to atomistic simulation, where evaluation requires multi-objective constraints with hard physical-feasibility gates across HPC experiments.

\textbf{Automated optimization.}
Bayesian methods (Optuna~\cite{akiba_optuna_2019}, BOHB~\cite{falkner_bohb_2018}), neural architecture search~\cite{zoph_nas_2017,elsken_nas_survey_2019}, and population-based training~\cite{jaderberg_pbt_2017} operate over predefined parameter spaces with scalar objectives.
MLIPilot differs in three respects: agents make arbitrary code changes (architectures, loss functions), the objective includes hard feasibility gates, and agents reason causally about why changes succeed or fail, accumulating knowledge across iterations in a manner more akin to human researchers than black-box optimizers.

\section{Method}

\subsection{System Architecture}

MLIPilot operates as a closed loop (Algorithm~\ref{alg:loop}): the LLM agent proposes a code change, the system trains a model on an HPC cluster, evaluates the result against the scorecard, and either accepts the improvement or reverts to the previous best.
This cycle repeats until the iteration budget is exhausted or the agent converges.
Five components implement this loop:

\begin{enumerate}[nosep,leftmargin=*]
    \item \textbf{Data Inspector}: Parses extXYZ datasets via ASE~\cite{hjorth_larsen_ase_2017}, identifies species, periodicity, and property keys, and generates train/valid/test splits (70/20/10) when absent.

    \item \textbf{Template Generator}: Synthesizes \texttt{train.py} with an editable experiment surface above a fixed evaluation harness (separated by a \texttt{\# FIXED HARNESS} sentinel), plus a scorecard with targets parsed from the user's goal prompt.

    \item \textbf{Agent Loop}: Orchestrates LLM tool-calling with retry logic, context truncation at 80k tokens, and dynamic early stopping via plateau detection.

    \item \textbf{HPC Executor}: Manages Slurm job lifecycle with exponential-backoff polling and local-GPU fallback for pre-allocated environments.

    \item \textbf{Scorecard Evaluator}: Computes composite score, enforces hard gates, and issues accept/reject decisions with automatic workspace revert on rejection.
\end{enumerate}

\subsection{Physically Constrained Scorecard}

A trained MLIP must satisfy multiple objectives simultaneously: low energy error, low force error, stable molecular dynamics, and sufficient throughput.
Rather than optimizing a single loss, MLIPilot evaluates each candidate model against a \emph{scorecard} that enforces two acceptance criteria:

\begin{enumerate}[nosep,leftmargin=*]
    \item \textbf{Improvement}: the candidate's composite score must be strictly better than the current best.
    \item \textbf{Physical feasibility}: \emph{every} metric must fall within a hard gate set at $4\times$ the user-specified target.
\end{enumerate}

\noindent Both conditions must hold for acceptance; otherwise the candidate is rejected and the workspace reverts to the previous best configuration.

\textbf{Why hard gates?}
A model with excellent energy accuracy but catastrophic NVE drift produces unphysical trajectories.
Composite scores alone cannot prevent this: a low energy penalty can mask a dangerously high drift.
Hard gates enforce that no single metric can be worse than $4\times$ target regardless of how favorable other metrics appear.

\textbf{Concrete example.}
On QM7, the energy target is 10~meV/atom, so the hard gate is 40~meV/atom.
The default baseline produces 51.98~meV/atom, which exceeds the 40~meV gate and is therefore rejected.
GPT-5.5's first optimization (energy\_weight~$=$~10) reduces energy to 21.89~meV/atom, passing all gates and achieving score 1.490 (accepted).
A subsequent experiment (500 epochs) passes the energy gate but causes NVE drift of 6.59~meV/atom/ps, exceeding the 4~meV/atom/ps drift gate, and is rejected despite lower energy.

\textbf{Formal definition.}
Each metric $m_i$ has a target $t_i$, weight $w_i$, direction $d_i$
(lower or higher is better), and penalty cap $c_i$. For metrics with
hard gates, the gate is set to $g_i = 4\,t_i$; as shown in
Table~\ref{tab:scorecard}, hard gates are not applied to all metrics.
The per-metric penalty ratio is
\begin{equation}
    p_i = \begin{cases}
        \min(x_i / t_i,\; c_i) & d_i = \text{lower-is-better} \\
        \min(t_i / x_i,\; c_i) & d_i = \text{higher-is-better}
    \end{cases}
\end{equation}
where $x_i$ is the measured value and the cap $c_i$ prevents any single metric from dominating the score.
The composite score is $S = \sum w_i p_i / \sum w_i$, where $S{=}1.0$ means all metrics exactly meet their targets and lower is better.
A candidate is accepted if:
\begin{equation}
    S_\text{new} < S_\text{best} \quad\text{and}\quad x_i \leq g_i \;\;\forall\, i
\end{equation}

\textbf{Adaptive targets.}
Targets are parsed automatically from natural-language goal prompts (e.g., ``energy $<$ 10~meV/atom'' $\to$ $t_\text{energy}{=}10$, $g_\text{energy}{=}40$), enabling users to specify dataset-appropriate constraints in a single command without manual scorecard configuration.

\subsection{Tool Interface}

Agents interact through six typed tools: \texttt{read\_file}, \texttt{write\_file}, \texttt{edit\_file}, \texttt{list\_files}, \texttt{submit\_and\_wait}, and \texttt{finish}.
Only \texttt{train.py} above the harness sentinel is writable.
\texttt{submit\_and\_wait} requires three structured arguments: \texttt{hypothesis} (what should improve), \texttt{target\_metric} (which metric), and \texttt{risk} (what might degrade), enforcing scientific discipline and producing auditable experiment records.

\subsection{Integrity Enforcement}

SHA-256 hashes of the evaluation harness and scorecard are computed at initialization and verified before every submission.
This prevents the agent from inflating scores by modifying test data, evaluation code, or scoring criteria.

\begin{algorithm}[t]
\caption{MLIPilot Agent Loop}
\label{alg:loop}
\small
\begin{algorithmic}[1]
\REQUIRE Dataset $D$, goal $G$, budget $B$, LLM $M$
\STATE $\sigma \leftarrow$ \textsc{ParseScorecard}($G$)
\STATE $\tau \leftarrow$ \textsc{GenerateCode}($D$, $\sigma$)
\STATE $h_0 \leftarrow$ SHA256(harness $\|$ $\sigma$)
\STATE $S^* \leftarrow \infty$; $n \leftarrow 0$
\WHILE{$n < B$ \AND \NOT converged}
    \STATE resp $\leftarrow M$.\textsc{Chat}(context, tools)
    \FOR{tc $\in$ resp.tool\_calls}
        \IF{tc = \texttt{submit\_and\_wait}}
            \STATE \textbf{verify} SHA256 $= h_0$
            \STATE metrics $\leftarrow$ \textsc{TrainAndEvaluate}($\tau$)
            \STATE $S, \text{pass} \leftarrow$ \textsc{Score}(metrics, $\sigma$)
            \STATE $n \leftarrow n + 1$
            \IF{$S < S^*$ \AND pass}
                \STATE $S^* \leftarrow S$; \textsc{Snapshot}()
            \ELSE
                \STATE \textsc{Revert}()
            \ENDIF
        \ENDIF
    \ENDFOR
\ENDWHILE
\end{algorithmic}
\end{algorithm}

\subsection{MACE Backend}

Training invokes \texttt{mace\_run\_train}~\cite{batatia_mace_2022}.
The fixed evaluation harness benchmarks each trained model on four axes: (i)~energy/force/stress MAE on held-out test frames, (ii)~throughput via 10 perturbed evaluations, (iii)~geometry relaxation (FIRE, $f_\text{max}{=}0.05$~eV/\AA), and (iv)~NVE MD stability (300~K, 0.5~fs timestep, 5,000~steps in total, drift from linear regression of total energy vs.\ time). Although the experiments reported here focus on MACE, MLIPilot implements a backend registry designed to support additional MLIP frameworks, including NequIP, Allegro, CHGNet, and other architectures.

\section{Experiments}

\subsection{Datasets}

\textbf{QM7 B3LYP}: Starting from the 7,165 organic molecules (C, H, N, O, S) in original QM7 daatset~\cite{rupp_qm7_2012,blum_qm7_gdb_2009}, we
generated B3LYP/6-31G(d) single-point energies and forces and split the
resulting dataset into 5,015/1,433/717 train/validation/test structures. 7,165 organic molecules (C, H, N, O, S) with B3LYP/6-31G(d) energies and forces.
Split 5,015/1,433/717.
Non-periodic, chemically diverse, with large variance in molecular size and energy scale.
This dataset presents a particular challenge for MLIPs because the wide energy distribution across different molecular compositions rewards models that can robustly handle statistical outliers.

\textbf{Cu EMT}: 200 strained/rattled FCC Cu supercells (32~atoms, $2{\times}2{\times}2$) with energies, forces, and Voigt stress from ASE's Effective Medium Potential (EMT) calculator.
The 200 structures were split into 140 training, 40 validation, and 20 test structures. The dataset is periodic and chemically simple, but geometrically diverse,
making it a useful benchmark for testing whether agents can drive models
toward tight convergence targets, including sub-meV/atom energy errors
relative to EMT reference labels.

\subsection{Protocol}

Table~\ref{tab:scorecard} defines scorecard parameters.
Four agents were benchmarked: GPT-5.5 (OpenAI reasoning model), GPT-4.1 (OpenAI tool-optimized), Mistral-24B (self-hosted via vLLM on H100), and Qwen3-32B (self-hosted via vLLM on H100).
Each received budget $B{=}10$ (up to ten train-and-evaluate iterations), identical MACE defaults, and a natural-language goal prompt specifying numeric targets.
Because scorecard feedback is used throughout the agent
loop, we treat the repeatedly evaluated held-out split as a
validation/selection set rather than a fully blind test set; a separate
sealed test set will be required for definitive generalization estimates.

\begin{table}[t]
\centering
\caption{Scorecard configuration. Targets parsed from goal; gates at $4\times$ target. Perfect score $S{=}1.0$.}
\label{tab:scorecard}
\small
\begin{tabular}{@{}lcccccc@{}}
\toprule
Metric & Dir & \multicolumn{2}{c}{Target} & $w$ & Cap & Gate \\
\cmidrule(lr){3-4}
 & & QM7 & Cu & & & ($4t$) \\
\midrule
E (meV/atom) & $\downarrow$ & 10 & 1 & 3 & 10 & 40/4 \\
F (meV/\AA) & $\downarrow$ & 10 & 5 & 3 & 10 & 40/20 \\
Thr.\ (steps/s) & $\uparrow$ & 50 & 50 & 0.3 & 5 & -- \\
Drift (meV/atom/ps) & $\downarrow$ & 1.0 & 0.5 & 1 & 10 & 4/2 \\
$\sigma$ (GPa) & $\downarrow$ & -- & 0.5 & 0.5 & 10 & --/2 \\
\bottomrule
\end{tabular}
\end{table}

\subsection{QM7 Results}

\begin{table}[t]
\centering
\caption{QM7 B3LYP final results. All baselines fail hard gates ($S{\approx}2.6$, E${}>40$~meV gate). B=budget exhausted, S=LLM stopped.}
\label{tab:qm7-results}
\small
\begin{tabular}{@{}lccccccl@{}}
\toprule
Model & Score & E & F & Drift & Thr. & Tokens & Stop \\
 & $\downarrow$ & meV/atom & meV/\AA & meV/atom/ps & steps/s & & \\
\midrule
GPT-5.5 & \textbf{0.831} & \textbf{9.52} & \textbf{9.83} & \textbf{0.051} & 70 & 244k & B \\
Mistral & 1.061 & 14.32 & 9.94 & 0.111 & 42 & 256k & B \\
GPT-4.1 & 1.288 & 17.83 & 12.17 & 0.082 & 47 & 205k & B \\
Qwen3 & 1.818 & 28.71 & 11.58 & 0.938 & 61 & 486k & S \\
\bottomrule
\end{tabular}
\end{table}

\begin{figure}[t]
\centering
\includegraphics[width=\columnwidth]{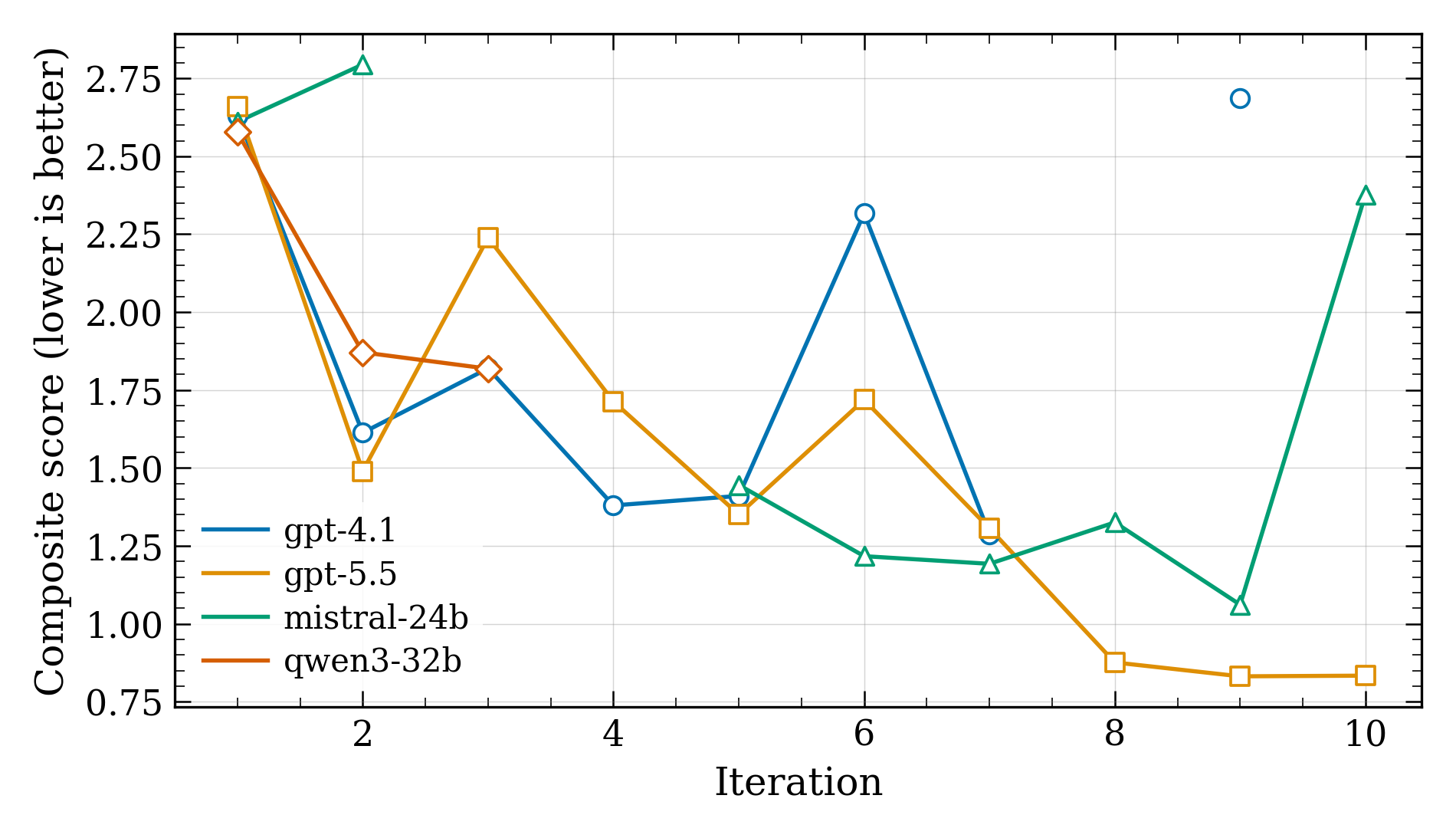}
\caption{Composite score convergence on QM7 (lower is better). All agents start from infeasible baselines ($S{\approx}2.6$, hard gates violated). Missing points indicate failed jobs that produced no score: Mistral-24B iters 3--4 (code errors in LR scheduler), GPT-4.1 iter 8 (failed with correlation=4). GPT-5.5 achieves the steepest descent through architectural innovation at iterations 5 and 7.}
\label{fig:qm7-convergence}
\end{figure}

The QM7 benchmark reveals the clearest separation between agents that reason scientifically and those that tune parameters.
All agents begin from baselines that \emph{violate hard gates}: the default model produces 51.98~meV/atom energy MAE, exceeding the 40~meV gate (Table~\ref{tab:gpt55-qm7}, iteration~1: rejected).
The baseline force MAE (10.42~meV/\AA) is already near target, so the primary challenge is reducing energy error without destabilizing dynamics.

\textbf{GPT-5.5} is the only agent to make \emph{architectural} changes.
Table~\ref{tab:gpt55-qm7} shows its full trajectory.
After establishing feasibility via energy weighting (iter~2, score 1.49), it attempts longer training (iter~3), which \emph{passes} the energy gate but causes catastrophic NVE drift (6.59~meV/atom/ps, above the 4.0~meV/atom/ps hard gate), illustrating the failure mode hard gates are designed to catch.
Rather than retrying training duration, GPT-5.5 pivots to architecture: switching to ScaleShiftMACE (iter~5, score 1.35$\to$accepted), hypothesizing that explicit output normalization suits the wide energy distribution in QM7's diverse molecular compositions.
It then substitutes Huber loss for squared error (iter~7), reasoning about robustness to outlier configurations, which drops force MAE from 12.52 to 9.61~meV/\AA.
The final configuration (ScaleShiftMACE + Huber + energy\_weight~30 + 300~epochs) satisfies both accuracy targets, achieving energy MAE below
10~meV/atom and force MAE below 10~meV/\AA{}, with a low NVE drift of only 0.051~meV/atom/ps.

\begin{table*}[t]
\centering
\caption{GPT-5.5 on QM7: autonomous discovery of ScaleShiftMACE (iter~5) and Huber loss (iter~7). These are not hyperparameter adjustments but changes to the model's functional form and training objective. Best (iter~9, bold) achieves both targets.}
\label{tab:gpt55-qm7}
\footnotesize
\begin{tabular}{@{}clcccccl@{}}
\toprule
\# & Intervention & E (meV/at) & F (meV/\AA) & Drift & Thr. & Score & Decision \\
 & & & & (meV/at/ps) & (steps/s) & & \\
\midrule
1 & Default baseline & 51.98 & 10.42 & 0.469 & 68 & 2.659 [\xmark] & Rej (gate) \\
2 & energy\_weight $= 10$ & 21.89 & 13.01 & 0.183 & 68 & 1.490 & Acc \\
3 & 500 epochs & 19.62 & 12.13 & 6.594 & 67 & 2.239 [\xmark] & Rej (gate) \\
4 & lr $= 0.001$ & 24.80 & 13.77 & 0.714 & 68 & 1.713 & Rej (score) \\
5 & \textit{ScaleShiftMACE} & 17.11 & 12.52 & 0.756 & 70 & 1.351 & Acc \\
6 & 3 interactions & 18.64 & 13.85 & 2.486 & 48 & 1.719 & Rej (score) \\
7 & \textit{Huber loss} & 19.18 & 9.61 & 0.684 & 70 & 1.306 & Acc \\
8 & ew $= 30$ & 10.48 & 10.05 & 0.020 & 70 & 0.876 & Acc \\
\textbf{9} & \textbf{300 ep + ew30 + Huber + SS} & \textbf{9.52} & \textbf{9.83} & \textbf{0.051} & \textbf{70} & \textbf{0.831} & \textbf{Acc} \\
10 & 400 epochs & 8.70 & 9.47 & 0.418 & 70 & 0.833 & Rej (score) \\
\bottomrule
\end{tabular}
\end{table*}

\begin{figure}[t]
\centering
\includegraphics[width=\columnwidth]{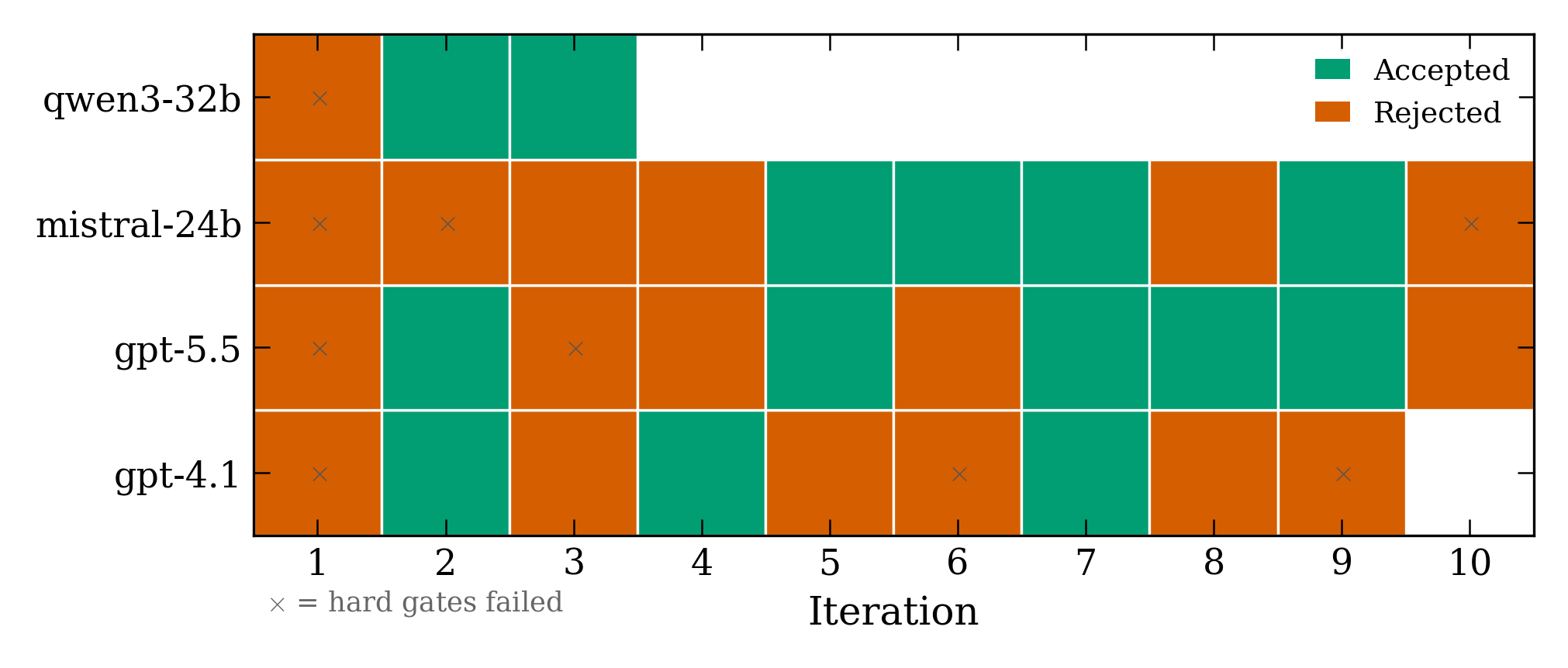}
\caption{Accept/reject decisions on QM7. Green: accepted; red: rejected; $\times$: hard gates failed. GPT-5.5's iterations 3 and baseline fail gates, while all other rejections are due to non-improving scores. Mistral-24B has 2 FAILED jobs (iters 3--4) before recovering.}
\label{fig:qm7-decisions}
\end{figure}

\textbf{Mistral-24B} achieves the second-best score (1.061), surpassing GPT-4.1 despite being a smaller open-weight model.
Its strategy is instructive: after two failed jobs (iterations 3--4, code errors with LR schedulers), it recovers and pursues a capacity-then-duration approach: increasing to 256~irreps with 3~interactions (iter~6, score 1.217), then extending training to 1000~epochs (iter~9, score 1.061).
While lacking GPT-5.5's architectural innovation, Mistral compensates through persistent exploration and willingness to invest heavily in training duration (4.9~hours for the final 1000-epoch run vs.\ GPT-5.5's 38~minutes for 300~epochs).
Mistral-24B and Qwen3-32B ran on a pre-allocated local GPU, where training executes synchronously.
Its force MAE (9.94~meV/\AA) is comparable to GPT-5.5's (9.83), but its energy MAE remains 50\% higher (14.32 vs.\ 9.52) without the ScaleShift normalization.

\textbf{GPT-4.1} achieved score 1.288 through parametric optimization (energy\_weight~10, 256~irreps, 3~interactions).
Its attempt to increase many-body correlation from 3 to 4 (iter~8) failed and produced no score (visible as a missing point in Figure~\ref{fig:qm7-convergence}).
A subsequent LR scheduler experiment (iter~9) produced catastrophic NVE drift, a failure invisible to energy/force MAE that hard gates correctly reject.

\textbf{Qwen3-32B} consumed 486k tokens in only 3 iterations (162k/iteration with 20 edit calls per 3 submissions) before the LLM stopped responding.
It achieved 1.818 through simple weight adjustments, never attempting architectural or training-duration changes.

\begin{figure}[t]
\centering
\includegraphics[width=\columnwidth]{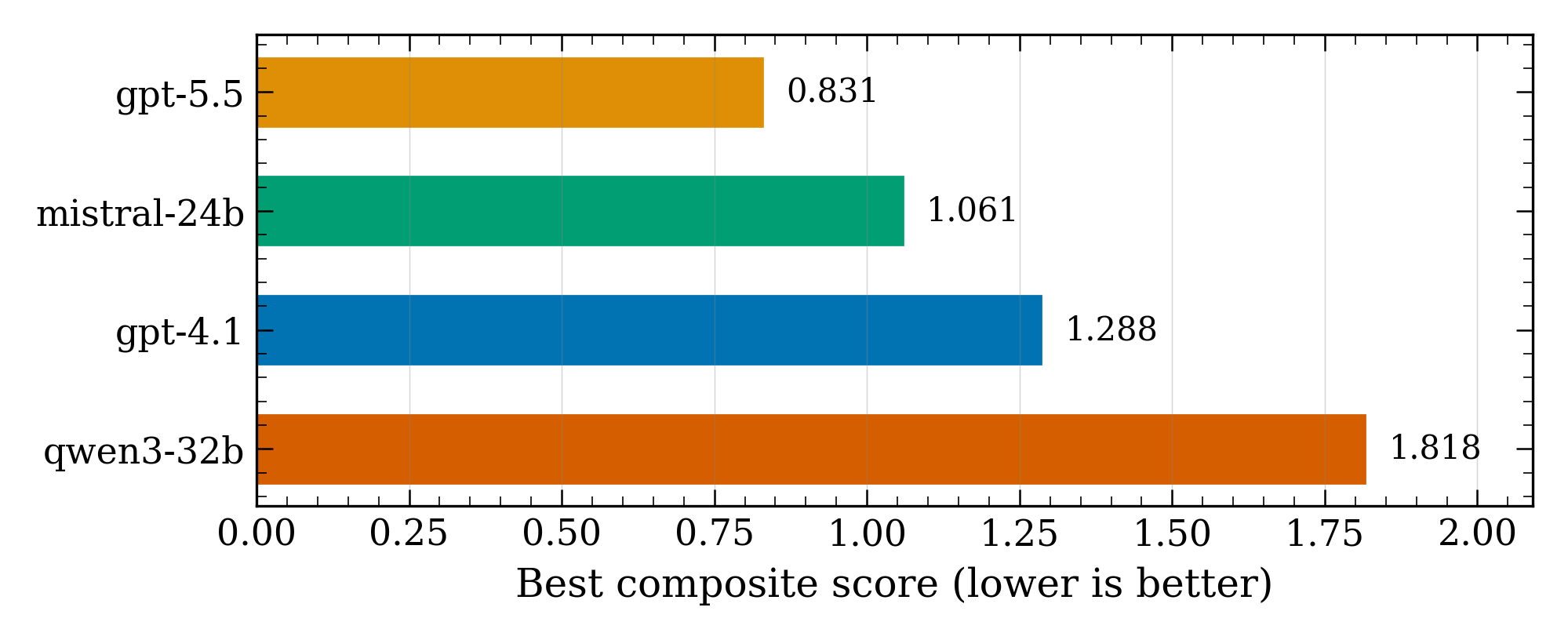}
\caption{Final score ranking on QM7. GPT-5.5 achieves $1.3\times$ better score than Mistral-24B and $1.5\times$ better than GPT-4.1.}
\label{fig:qm7-ranking}
\end{figure}

\clearpage
\subsection{Cu EMT Results}

\begin{table}[t]
\centering
\caption{Cu EMT final results. All baselines fail hard gates ($S{\approx}4.5$, E${}>4$~meV gate). B=budget exhausted, C=converged.}
\label{tab:cu-emt-results}
\small
\begin{tabular}{@{}lcccccc@{}}
\toprule
Model & Score & E & F & $\sigma$ & Drift & Stop \\
 & $\downarrow$ & meV/atom & meV/\AA & GPa & meV/atom/ps & \\
\midrule
GPT-5.5 & \textbf{0.401} & \textbf{0.57} & \textbf{1.40} & \textbf{0.001} & 0.121 & B \\
GPT-4.1 & 0.653 & 1.06 & 2.45 & 0.001 & 0.110 & B \\
Mistral & 1.190 & 1.93 & 4.91 & 0.003 & \textbf{0.097} & B \\
Qwen3 & 1.301 & 2.37 & 4.09 & 0.003 & 0.108 & C \\
\bottomrule
\end{tabular}
\end{table}

\begin{figure}[t]
\centering
\includegraphics[width=\columnwidth]{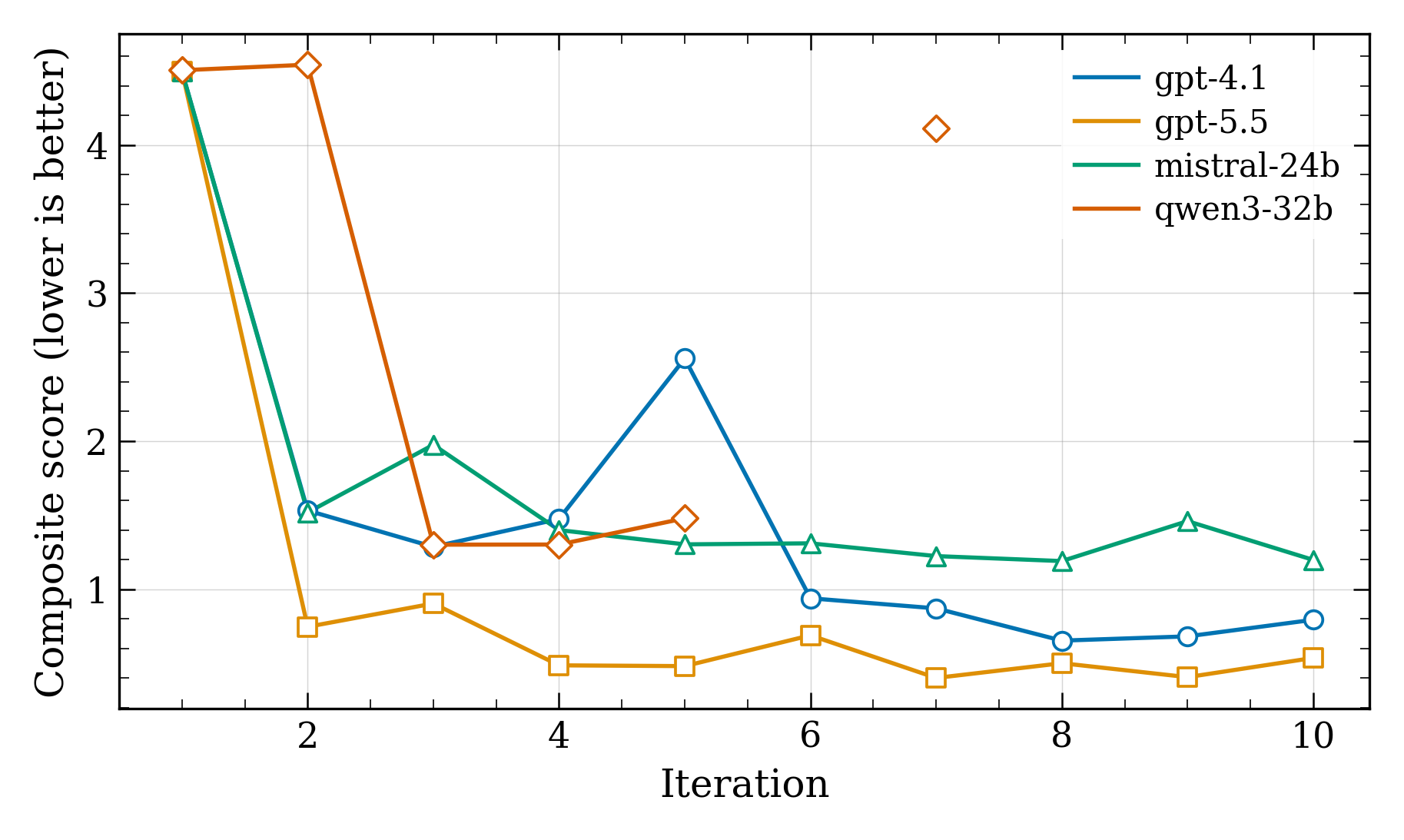}
\caption{Composite score convergence on Cu EMT. Missing point: Qwen3-32B iter~6 (FAILED job from invalid loss function string). GPT-5.5 achieves
the best final score through progressive epoch scaling, reaching 0.401 at iteration~7.}
\label{fig:convergence}
\end{figure}

All agents begin from baselines that \emph{violate hard gates}: the default model produces 12.7~meV/atom energy MAE, which exceeds the 4~meV gate (Table~\ref{tab:gpt55-cu}, iteration~1: rejected).
The agent must first make a large enough change to cross the feasibility boundary before any configuration can be accepted.

Table~\ref{tab:gpt55-cu} details GPT-5.5's full trajectory.
The agent's first change (increasing from 50 to 500 training epochs) drops energy MAE from 12.7 to 1.22~meV/atom, passing all hard gates for the first time and achieving score 0.746 (accepted).
It then tries increasing energy loss weight (iteration~3), which passes gates but produces a \emph{worse} score (0.904 vs.\ 0.746), so it is rejected.
The agent learns from this and returns to epoch scaling: 1000~epochs (accepted, 0.486) then 2000~epochs (accepted, 0.481).
At iteration~7, it combines 2000~epochs with a third message-passing interaction layer, achieving 0.57~meV/atom energy MAE (well below the 1~meV target) at the cost of reduced throughput (46 vs.\ 65~steps/s).
Three subsequent attempts (smaller model, more epochs, different seed) all produce valid models that pass gates, but none beat the current best score of 0.401, so all are rejected.

\begin{table*}[t]
\centering
\caption{GPT-5.5 iteration history on Cu EMT. The progressive epoch scaling (50$\to$500$\to$1000$\to$2000) is an emergent strategy. Best iteration (7, bold) achieves sub-meV accuracy by adding a third interaction layer.}
\label{tab:gpt55-cu}
\footnotesize
\begin{tabular}{@{}clcccccccl@{}}
\toprule
\# & Intervention & E (meV/atom) & F (meV/\AA) & $\sigma$ (GPa) & Drift & Thr. & Score & Gates & Dec \\
\midrule
1 & Default (50 epochs) & 12.69 & 7.70 & 0.016 & 0.112 & 65 & 4.499 & \xmark & Rej \\
2 & 500 epochs & 1.22 & 2.85 & 0.002 & 0.104 & 65 & 0.746 & \cmark & Acc \\
3 & energy\_weight $= 10$ & 1.67 & 2.69 & 0.002 & 0.103 & 65 & 0.904 & \cmark & Rej \\
4 & 1000 epochs & 0.69 & 2.13 & 0.001 & 0.107 & 66 & 0.486 & \cmark & Acc \\
5 & 2000 epochs & 0.77 & 1.76 & 0.001 & 0.069 & 65 & 0.481 & \cmark & Acc \\
6 & lr $= 0.005$ & 1.30 & 1.75 & 0.001 & 0.088 & 65 & 0.689 & \cmark & Rej \\
\textbf{7} & \textbf{3 interactions} & \textbf{0.57} & \textbf{1.40} & \textbf{0.001} & \textbf{0.121} & \textbf{46} & \textbf{0.401} & \cmark & \textbf{Acc} \\
8 & Narrower (64 irreps) & 0.80 & 1.61 & 0.001 & 0.113 & 46 & 0.501 & \cmark & Rej \\
9 & 3000 epochs & 0.59 & 1.32 & 0.001 & 0.145 & 46 & 0.406 & \cmark & Rej \\
10 & Seed change & 0.89 & 1.54 & 0.001 & 0.128 & 46 & 0.536 & \cmark & Rej \\
\bottomrule
\end{tabular}
\end{table*}

\begin{figure}[t]
\centering
\includegraphics[width=\columnwidth]{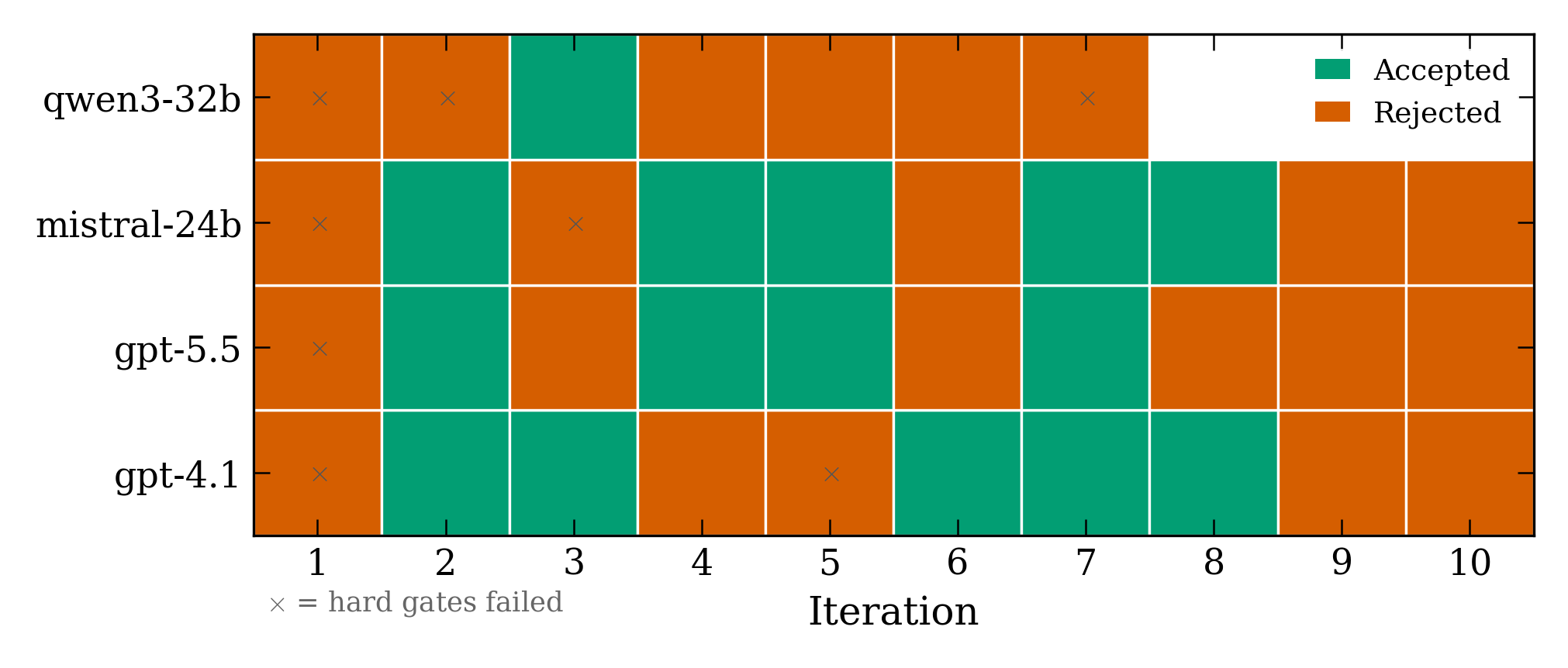}
\caption{Accept/reject decisions on Cu EMT. Green: accepted; red: rejected; $\times$: hard gates failed. All baselines (column~1) fail gates. GPT-5.5 requires only one iteration to reach feasibility.}
\label{fig:decisions}
\end{figure}

\textbf{GPT-4.1} adopted a capacity-first strategy: increasing irreps to 256 (iter~3), then energy weighting (ew$=$30, iter~7), then extended training (500~epochs, iter~8).
It reached score 0.653 with E$=$1.06~meV/atom, just above the 1~meV target.

\textbf{Mistral-24B} made 5 accepted steps through methodical single-variable changes (epochs, weighting, interactions, irreps, LR), achieving 1.190 but never discovering that training duration beyond 200~epochs is the dominant optimization axis.

\textbf{Qwen3-32B} converged early at score 1.301 after 7~iterations (3 consecutive rejections triggered early stopping).
Its single acceptance (iter~3: 200 epochs + LR scheduler) was the only configuration that passed gates.
Subsequent attempts included capacity increase (rejected, worse score), energy weighting (rejected, marginal), changing the loss function string to an invalid value (iter~6: job FAILED, visible as a missing point in Figure~\ref{fig:convergence}), and a corrected loss variant that failed hard gates.

\begin{figure}[t]
\centering
\includegraphics[width=\columnwidth]{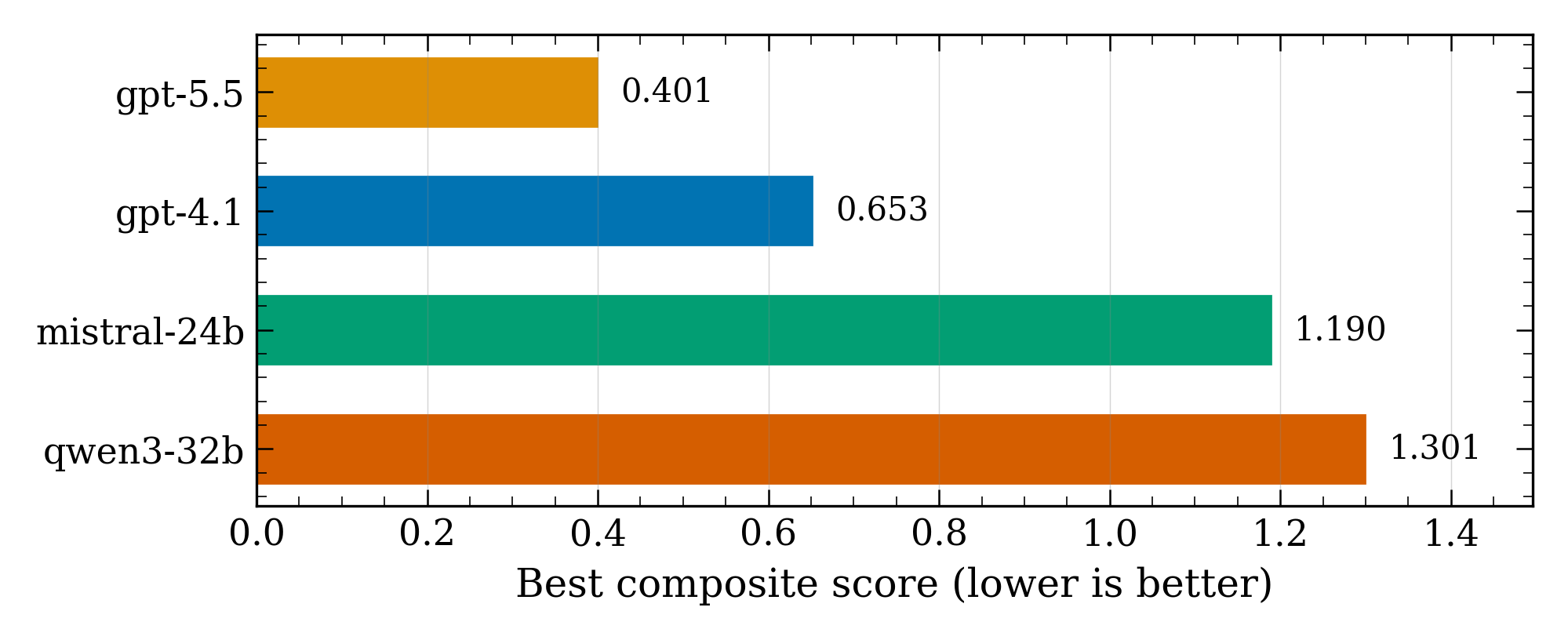}
\caption{Final score ranking on Cu EMT. GPT-5.5 achieves $1.6\times$ lower score than GPT-4.1 and $3.2\times$ lower than Qwen3-32B.}
\label{fig:ranking}
\end{figure}

\clearpage
\subsection{Cross-Dataset Analysis}

\begin{table}[t]
\centering
\caption{Efficiency summary. Improvement $= S_\text{baseline}/S_\text{best}$.}
\label{tab:cost}
\small
\begin{tabular}{@{}llccccc@{}}
\toprule
Model & Data & Iters & Tokens & Impr. & Acceptance\ \% \\
\midrule
GPT-5.5 & QM7 & 10 & 244k & 3.2$\times$ & 50 \\
Mistral & QM7 & 10 & 256k & 2.5$\times$ & 40 \\
GPT-4.1 & QM7 & 9 & 205k & 2.0$\times$ & 33 \\
Qwen3 & QM7 & 3 & 486k & 1.4$\times$ & 67 \\
\midrule
GPT-5.5 & Cu EMT & 10 & 546k & 11.2$\times$ & 40 \\
GPT-4.1 & Cu EMT & 10 & 183k & 6.9$\times$ & 50 \\
Mistral & Cu EMT & 10 & 232k & 3.8$\times$ & 50 \\
Qwen3 & Cu EMT & 7 & 184k & 3.5$\times$ & 14 \\
\bottomrule
\end{tabular}
\end{table}

\begin{figure}[t]
\centering
\includegraphics[width=\columnwidth]{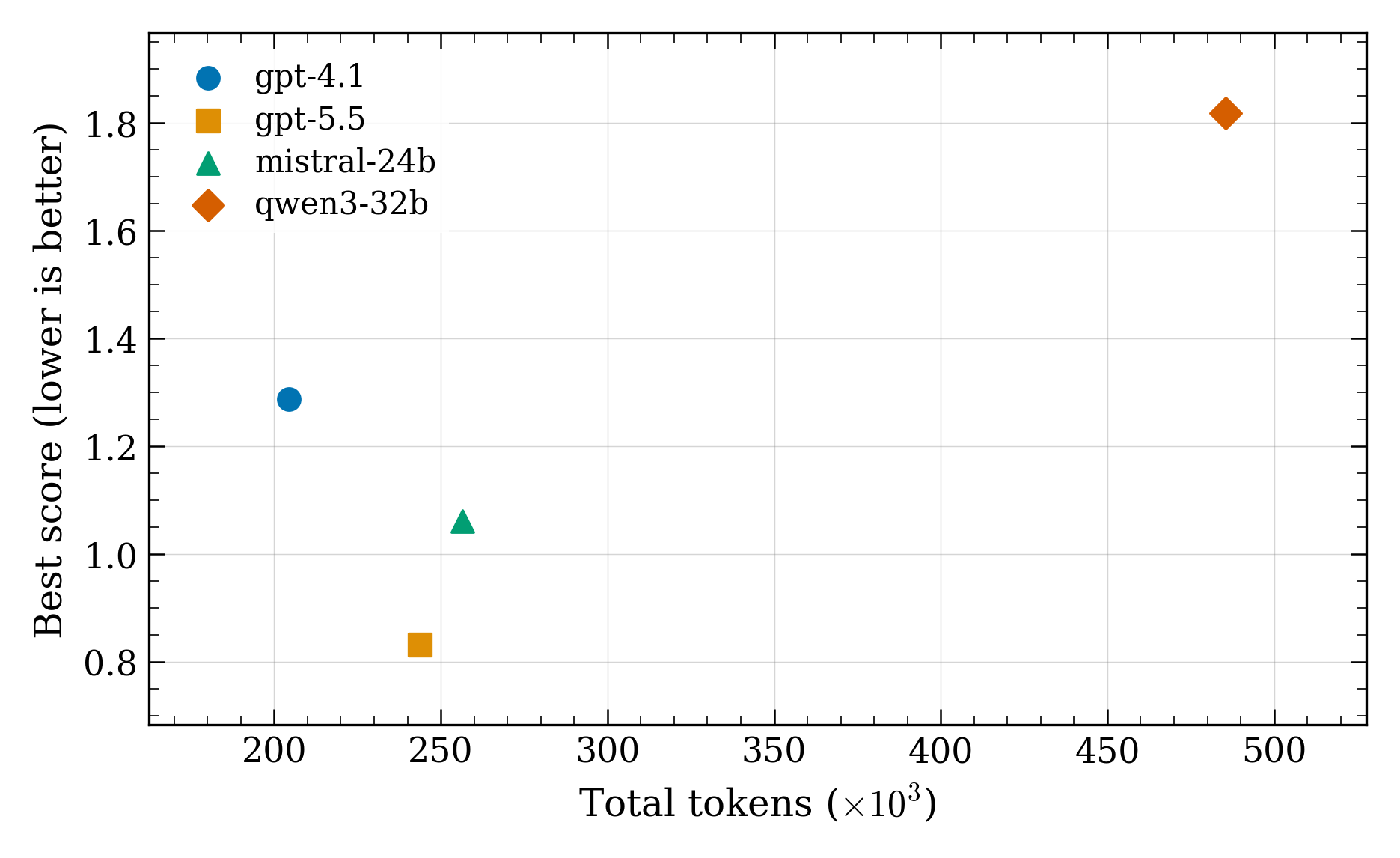}
\caption{Token expenditure vs.\ final score (QM7). Qwen3-32B's 486k tokens (162k/iter) yield only 1.4$\times$ improvement, while GPT-5.5's 244k produce 3.2$\times$.}
\label{fig:qm7-cost}
\end{figure}

Four patterns emerge from Table~\ref{tab:cost}:

(1)~\textbf{Reasoning quality dominates scale.}
GPT-5.5 achieves 3.2--11.2$\times$ improvement through qualitatively different interventions (architecture, loss function); GPT-4.1 achieves 2.0--6.9$\times$ through parametric tuning; open-weight models 1.4--3.8$\times$ through incremental changes.

(2)~\textbf{Token count does not predict quality.}
Qwen3-32B's 486k tokens on QM7 (162k/iteration with excessive edit calls) yield only 1.4$\times$ improvement, while GPT-5.5's 244k yield 3.2$\times$.
On Cu EMT, GPT-5.5's 546k tokens produce 11.2$\times$ vs.\ Mistral's 232k for 3.8$\times$.

(3)~\textbf{Open-weight models can outperform proprietary ones.}
Mistral-24B surpasses GPT-4.1 on QM7 (2.5$\times$ vs.\ 2.0$\times$ improvement) through willingness to invest in extended training (1000 epochs).
This suggests that some optimization landscapes reward persistence over sophistication, as long as the agent can identify and execute a viable strategy.

(4)~\textbf{Tighter targets amplify differentiation.}
Cu EMT improvement factors span 3.5--11.2$\times$ (3.2$\times$ dynamic range) vs.\ QM7's 1.4--3.2$\times$ (2.3$\times$ range), reflecting that sub-meV targets require deeper reasoning to satisfy.

\section{Discussion}

\subsection{Emergent Scientific Behaviors}

Without explicit instruction, agents develop behaviors characteristic of expert researchers:

\begin{itemize}[nosep,leftmargin=*]
    \item \textbf{Progressive scaling}: GPT-5.5 escalates epochs 50$\to$500$\to$1000$\to$2000 on Cu~EMT, monotonically increasing computational investment as diminishing returns require deeper optimization.
    \item \textbf{Architecture switching}: Mid-campaign on QM7, GPT-5.5 replaces MACE with ScaleShiftMACE, hypothesizing that molecular energies benefit from explicit output normalization. This is a change to the model's functional form, not a hyperparameter.
    \item \textbf{Loss innovation}: GPT-5.5 replaces squared-error loss with Huber loss on QM7, reasoning about robustness to outlier configurations in a chemically diverse dataset. Force MAE drops 23\% (12.52$\to$9.61~meV/\AA).
    \item \textbf{Multi-level diagnosis}: When 500-epoch training caused drift failure on QM7, GPT-5.5 sequentially tried LR reduction (insufficient), then pivoted to an architectural change (successful).
    \item \textbf{Error recovery}: Mistral-24B on QM7 encountered two consecutive FAILED jobs (LR scheduler syntax errors) but recovered by fixing the code and continuing optimization.
    \item \textbf{Risk-aware retreat}: After LR reduction worsened Cu~EMT score, GPT-5.5 abandoned the LR dimension entirely and explored model depth instead.
\end{itemize}

\subsection{The Role of Hard Gates}

Hard gates at $4\times$ target serve three functions.
First, they exclude dynamically unstable models: every baseline and several optimized configurations (GPT-5.5 QM7 iter~3: drift 6.59; GPT-4.1 QM7 iter~9: catastrophic drift from LR scheduler; Mistral QM7 iter~10: drift from aggressive loss weighting) are correctly rejected despite acceptable energy/force MAE.
Second, they create a feasibility-first optimization landscape where agents must solve a constraint-satisfaction problem before standard minimization begins.
Third, the $4\times$ multiplier provides a reasonable heuristic boundary: potentials within $4\times$ of target retain qualitative correctness for screening applications, while those beyond it produce unphysical trajectories.

\subsection{What Distinguishes Effective Agent Researchers}

Optimization success correlates with three capabilities:
(1)~\emph{Creative search}: making qualitative changes (architecture, loss function) rather than only quantitative adjustments;
(2)~\emph{Compositional reasoning}: combining independently beneficial changes (ScaleShift + Huber + energy\_weight + epochs) into a jointly superior configuration;
(3)~\emph{Causal interpretation}: diagnosing \emph{why} a change failed (drift instability from overtraining) and pivoting to an orthogonal intervention rather than repeating similar modifications.

These capabilities are present in GPT-5.5, partially in GPT-4.1 and Mistral-24B, and largely absent in Qwen3-32B.
The ranking on QM7 (GPT-5.5 $>$ Mistral $>$ GPT-4.1 $>$ Qwen3) is particularly informative: Mistral's success over GPT-4.1 demonstrates that willingness to fully exhaust a working strategy (extended training) can compensate for lack of architectural innovation on datasets where training duration is a dominant axis.

\section{Conclusion}

MLIPilot demonstrates that tool-calling LLMs can function as autonomous interatomic-potential researchers, achieving 3.2--11.2$\times$ improvement in composite score and up to 22$\times$ reduction in energy MAE (12.69$\to$0.57~meV/atom on Cu EMT) within 10 iterations.
The physically constrained scorecard with $4\times$ hard gates is essential: all baselines fail gates, and configurations with catastrophic dynamics (drift $>$6~meV/atom/ps on QM7, failing the 4~meV/atom/ps gate) are correctly rejected despite composite-score improvement.
GPT-5.5's autonomous discovery of ScaleShiftMACE with Huber loss on QM7 represents a qualitative advance beyond hyperparameter optimization, demonstrating genuine scientific reasoning about which model variant and loss function suit a dataset's statistical structure.
Mistral-24B's ability to surpass GPT-4.1 on QM7 through persistent training-duration exploration suggests that the landscape of ML research tasks is multidimensional: some reward creative insight while others reward methodical exhaustion of viable strategies.

As language models advance in causal and compositional reasoning, systems like MLIPilot point toward a future where the bottleneck in atomistic simulation shifts from \emph{how to train} potentials to \emph{what physical questions to ask}, freeing domain scientists from the engineering of training pipelines.

\section*{Author Contributions}

E.O., S.A., and D.R.\ conceptualized the project.
E.O.\ developed the software, designed and executed all LLM benchmarks. E.O. and D.R. wrote the initial manuscript. S.A. and S.Z. validated the software. E.O. and K.P. generated the data for this work. All authors contributed to scientific reasoning.
All authors revised and approved the final manuscript.

\section*{Acknowledgments}

The authors thank the broader PsiQuantum application team. Authors particularly acknowledge helpful discussions with Arpan Kundu, Swagata Roy, and Arvin Kakekhani.

\bibliographystyle{unsrt}
\bibliography{references}

@article{batatia_mace_2022,
  title={{MACE}: Higher Order Equivariant Message Passing Neural Networks for Fast and Accurate Force Fields},
  author={Batatia, Ilyes and Kov{\'a}cs, D{\'a}vid P{\'e}ter and Simm, Gregor N C and Ortner, Christoph and Cs{\'a}nyi, G{\'a}bor},
  journal={Advances in Neural Information Processing Systems},
  volume={35},
  pages={11423--11436},
  year={2022}
}

@article{batzner_nequip_2022,
  title={{E(3)}-equivariant graph neural networks for data-efficient and accurate interatomic potentials},
  author={Batzner, Simon and Musaelian, Albert and Sun, Lixin and Geiger, Mario and Mailoa, Jonathan P and Kornbluth, Mordechai and Molinari, Nicola and Smidt, Tess E and Kozinsky, Boris},
  journal={Nature Communications},
  volume={13},
  pages={2453},
  year={2022}
}

@article{schutt_schnet_2018,
  title={{SchNet} -- A deep learning architecture for molecules and materials},
  author={Sch{\"u}tt, Kristof T and Sauceda, Huziel E and Kindermans, P-J and Tkatchenko, Alexandre and M{\"u}ller, Klaus-Robert},
  journal={The Journal of Chemical Physics},
  volume={148},
  pages={241722},
  year={2018}
}

@inproceedings{gasteiger_dimenetpp_2020,
  title={Fast and Uncertainty-Aware Directional Message Passing for Non-Equilibrium Molecules},
  author={Gasteiger, Johannes and Giri, Shankari and Margraf, Johannes T and G{\"u}nnemann, Stephan},
  booktitle={Machine Learning for Molecules Workshop, NeurIPS},
  year={2020}
}

@article{musaelian_allegro_2023,
  title={Learning local equivariant representations for large-scale atomistic dynamics},
  author={Musaelian, Albert and Batzner, Simon and Johansson, Anders and Sun, Lixin and Owen, Cameron J and Kornbluth, Mordechai and Kozinsky, Boris},
  journal={Nature Communications},
  volume={14},
  pages={579},
  year={2023}
}

@inproceedings{schutt_painn_2021,
  title={{PaiNN}: Polarizable Atom Interaction Neural Network},
  author={Sch{\"u}tt, Kristof T and Unke, Oliver T and Gastegger, Michael},
  booktitle={International Conference on Machine Learning},
  pages={9377--9388},
  year={2021}
}

@article{batatia_mace_mp_2024,
  title={A foundation model for atomistic materials chemistry},
  author={Batatia, Ilyes and Benner, Philipp and Chiang, Yuan and Elena, Alin M and Kov{\'a}cs, D{\'a}vid P and Riebesell, Janosh and others},
  journal={arXiv preprint arXiv:2401.00096},
  year={2024}
}

@article{chen_m3gnet_2022,
  title={A universal graph deep learning interatomic potential for the periodic table},
  author={Chen, Chi and Ong, Shyue Ping},
  journal={Nature Computational Science},
  volume={2},
  pages={718--728},
  year={2022}
}

@article{deng_chgnet_2023,
  title={{CHGNet} as a pretrained universal neural network potential for charge-informed atomistic modelling},
  author={Deng, Bowen and Zhong, Peichen and Jun, KyuJung and Riebesell, Janosh and Han, Kevin and Bartel, Christopher J and Ceder, Gerbrand},
  journal={Nature Machine Intelligence},
  volume={5},
  pages={1031--1041},
  year={2023}
}

@article{boiko_coscientist_2023,
  title={Autonomous chemical research with large language models},
  author={Boiko, Daniil A and MacKnight, Robert and Kline, Ben and Gomes, Gabe},
  journal={Nature},
  volume={624},
  pages={570--578},
  year={2023}
}

@article{bran_chemcrow_2024,
  title={{ChemCrow}: Augmenting large-language models with chemistry tools},
  author={Bran, Andres M and Cox, Sam and Schilter, Oliver and Baldassari, Carlo and White, Andrew D and Schwaller, Philippe},
  journal={Nature Machine Intelligence},
  volume={6},
  pages={525--535},
  year={2024}
}

@article{jablonka_llm_chemistry_2024,
  title={14 examples of how {LLMs} can transform materials science and chemistry: a reflection on a large language model hackathon},
  author={Jablonka, Kevin Maik and Ai, Qianxiang and Al-Feghali, Alexander and Baber, Shruti and Balcells, David and others},
  journal={Digital Discovery},
  volume={2},
  pages={1233--1250},
  year={2023}
}

@inproceedings{akiba_optuna_2019,
  title={{Optuna}: A Next-generation Hyperparameter Optimization Framework},
  author={Akiba, Takuya and Sano, Shotaro and Yanase, Toshihiko and Ohta, Takeru and Koyama, Masanori},
  booktitle={Proceedings of the 25th ACM SIGKDD International Conference on Knowledge Discovery \& Data Mining},
  pages={2623--2631},
  year={2019}
}

@inproceedings{falkner_bohb_2018,
  title={{BOHB}: Robust and Efficient Hyperparameter Optimization at Scale},
  author={Falkner, Stefan and Klein, Aaron and Hutter, Frank},
  booktitle={International Conference on Machine Learning},
  pages={1437--1446},
  year={2018}
}

@misc{karpathy_autoresearch,
  title={{AutoResearch}: {AI} Agents Running Research on Single-{GPU} Nanochat Training Automatically},
  author={Karpathy, Andrej},
  year={2026},
  howpublished={\url{https://github.com/karpathy/AutoResearch}},
  url={https://github.com/karpathy/AutoResearch}
}

@misc{karpathy_autoresearch_session,
  title={Nanochat: Autoresearch Round~1 Improvements},
  author={Karpathy, Andrej},
  year={2026},
  howpublished={\url{https://github.com/karpathy/nanochat/commit/6ed7d1d82cee16c2e26f45d559ad3338447a6c1b}},
  url={https://github.com/karpathy/nanochat/commit/6ed7d1d82cee16c2e26f45d559ad3338447a6c1b}
}

@article{hjorth_larsen_ase_2017,
  title={The Atomic Simulation Environment -- a {Python} library for working with atoms},
  author={Hjorth Larsen, Ask and Mortensen, Jens J{\o}rgen and Blomqvist, Jakob and Castelli, Ivano E and Christensen, Rune and Du{\l}ak, Marcin and Friis, Jesper and Groves, Michael N and Hammer, Bj{\o}rk and Hargus, Cory and others},
  journal={Journal of Physics: Condensed Matter},
  volume={29},
  pages={273002},
  year={2017}
}

@article{rupp_qm7_2012,
  title={Fast and accurate modeling of molecular atomization energies with machine learning},
  author={Rupp, Matthias and Tkatchenko, Alexandre and M{\"u}ller, Klaus-Robert and von Lilienfeld, O. Anatole},
  journal={Physical Review Letters},
  volume={108},
  pages={058301},
  year={2012}
}

@article{blum_qm7_gdb_2009,
  title={970 million druglike small molecules for virtual screening in the chemical universe database {GDB}-13},
  author={Blum, Lorenz C and Reymond, Jean-Louis},
  journal={Journal of the American Chemical Society},
  volume={131},
  pages={8732--8733},
  year={2009}
}

@inproceedings{brown_gpt3_2020,
  title={Language Models are Few-Shot Learners},
  author={Brown, Tom and Mann, Benjamin and Ryder, Nick and Subbiah, Melanie and Kaplan, Jared and Dhariwal, Prafulla and others},
  booktitle={Advances in Neural Information Processing Systems},
  volume={33},
  pages={1877--1901},
  year={2020}
}

@article{openai_gpt4_2023,
  title={{GPT-4} Technical Report},
  author={{OpenAI}},
  journal={arXiv preprint arXiv:2303.08774},
  year={2023}
}

@article{touvron_llama_2023,
  title={{LLaMA}: Open and Efficient Foundation Language Models},
  author={Touvron, Hugo and Lavril, Thibaut and Izacard, Gautier and Martinet, Xavier and Lachaux, Marie-Anne and others},
  journal={arXiv preprint arXiv:2302.13971},
  year={2023}
}

@book{hutter_automl_book_2019,
  title={Automated Machine Learning: Methods, Systems, Challenges},
  author={Hutter, Frank and Kotthoff, Lars and Vanschoren, Joaquin},
  publisher={Springer},
  year={2019}
}

@article{he_automl_survey_2021,
  title={{AutoML}: A Survey of the State-of-the-Art},
  author={He, Xin and Zhao, Kaiyong and Chu, Xiaowen},
  journal={Knowledge-Based Systems},
  volume={212},
  pages={106622},
  year={2021}
}

@article{behler_bpnn_2007,
  title={Generalized Neural-Network Representation of High-Dimensional Potential-Energy Surfaces},
  author={Behler, J{\"o}rg and Parrinello, Michele},
  journal={Physical Review Letters},
  volume={98},
  pages={146401},
  year={2007}
}

@inproceedings{gasteiger_gemnet_2021,
  title={{GemNet}: Universal Directional Graph Neural Networks for Molecules},
  author={Gasteiger, Johannes and Becker, Florian and G{\"u}nnemann, Stephan},
  booktitle={Advances in Neural Information Processing Systems},
  volume={34},
  pages={6790--6802},
  year={2021}
}

@article{fu_forces_are_not_enough_2023,
  title={Forces are not Enough: Benchmark and Critical Evaluation for Machine Learning Force Fields with Molecular Simulations},
  author={Fu, Xiang and Wu, Zhenghao and Wang, Wujie and Xie, Tian and Keten, Sinan and Gomez-Bombarelli, Rafael and Jaakkola, Tommi},
  journal={Transactions on Machine Learning Research},
  year={2023}
}

@article{schick_toolformer_2023,
  title={Toolformer: Language Models Can Teach Themselves to Use Tools},
  author={Schick, Timo and Dwivedi-Yu, Jane and Dess{\`i}, Roberto and Raileanu, Roberta and Lomeli, Maria and Hambro, Eric and Zettlemoyer, Luke and Cancedda, Nicola and Scialom, Thomas},
  journal={Advances in Neural Information Processing Systems},
  volume={36},
  year={2023}
}

@article{qin_tool_learning_2023,
  title={Tool Learning with Foundation Models},
  author={Qin, Yujia and Liang, Shengding and Ye, Yining and Zhu, Kunlun and Yan, Lan and Lu, Yaxi and others},
  journal={arXiv preprint arXiv:2304.08354},
  year={2023}
}

@article{romera_paredes_funsearch_2024,
  title={Mathematical discoveries from program search with large language models},
  author={Romera-Paredes, Bernardino and Barekatain, Mohammadamin and Novikov, Alexander and Balog, Matej and Kumar, M Pawan and Dupont, Emilien and Ruiz, Francisco J R and Ellenberg, Jordan S and Wang, Pengming and Fawzi, Omar and others},
  journal={Nature},
  volume={625},
  pages={468--475},
  year={2024}
}

@article{lu_ai_scientist_2024,
  title={The {AI} Scientist: Towards Fully Automated Open-Ended Scientific Discovery},
  author={Lu, Chris and Lu, Cong and Lange, Robert Tjarko and Foerster, Jakob and Clune, Jeff and Ha, David},
  journal={arXiv preprint arXiv:2408.06292},
  year={2024}
}

@inproceedings{zoph_nas_2017,
  title={Neural Architecture Search with Reinforcement Learning},
  author={Zoph, Barret and Le, Quoc V},
  booktitle={International Conference on Learning Representations},
  year={2017}
}

@article{elsken_nas_survey_2019,
  title={Neural Architecture Search: A Survey},
  author={Elsken, Thomas and Metzen, Jan Hendrik and Hutter, Frank},
  journal={Journal of Machine Learning Research},
  volume={20},
  pages={1--21},
  year={2019}
}

@article{jaderberg_pbt_2017,
  title={Population Based Training of Neural Networks},
  author={Jaderberg, Max and Dalibard, Valentin and Osindero, Simon and Czarnecki, Wojciech M and Donahue, Jeff and Razavi, Ali and Vinyals, Oriol and Green, Tim and Dunning, Iain and Simonyan, Karen and others},
  journal={arXiv preprint arXiv:1711.09846},
  year={2017}
}

\end{document}